\documentclass[prb,aps,showpacs,twocolumn,floatfix]{revtex4}

\usepackage{graphicx,color}
\usepackage{dcolumn}
\usepackage{bm}

\newcommand{\tonda}[1]{ \left( #1 \right) }
\newcommand{\graf}[1]{ \left\{ #1 \right\} }

\newcommand{\gr} [1]{\textbf #1}

\begin{document}

\title{\bf{The one-dimensional t-J model coupled to adiabatic phonons: A numerical investigation}}
\author{
Samuele Bissola$^{1}$
and Alberto Parola$^{1,2}$
}
\affiliation {
$^1$ Dipartimento di Fisica e Matematica, Universit\'a dell'Insubria, Via Valleggio 11
Como, Italy \\
$^2$ Istituto Nazionale per la Fisica della Materia, Como, Italy}

\date{\today}

\begin{abstract}
The ground state of the one-dimensional t-J model coupled with phonons in the adiabatic limit
is numerically investigated by use of the Lanczos technique at quarter filling.
Due to the interplay between the electron-electron Coulomb repulsion and electron-phonon interaction,
this model shows a sequence of lattice distortions leading to the formation of charge-density-waves 
and bond-order-waves. Moderate electron-electron and electron-lattice coupling may lead to 
coexistence of dimerization and tetramerization in the distortion pattern. Dimerization
leads to the formation of an ``antiferromagnetic"  Mott insulator, while tetramerization
gives rise to a spin-Peierls phase. By increasing the super-exchange coupling, antiferromagnetism
is inhibited due to the change of the distortion periodicity. 
\end{abstract}
\maketitle
\section{Introduction}
In the last decades the physical and chemical properties of quasi one-dimensional (1D)
systems have attracted growing attention due to the occurrence, in the same sample,
of several different phases, ranging from insulating to superconducting, 
when a control parameter, like pressure, is varied \cite{bourjer,wilhelm}.  
Inorganic compounds, as $CuGeO_3$\cite{hase} or
$NaV_2O_5$\cite{misterx}, show clear evidence in favor of the existence of
spin-Peierls states with broken lattice symmetry. 
Even more complex phase diagrams are displayed by quasi 1D organic materials: 
the $DI-$$DCNQI$ compounds \cite{meneghetti,dumm},
the Bechgaard salts series $(TMTSF)_2X$ and its sulfur analog (Fabre salts series) 
$(TMTTF)_2X$, where $X$ denotes different monovalent anions ($ X=PF_6,AsF_6,ClO_4,...$). 
These systems, which consist of stacks of weakly coupled chains, 
exhibit, at low temperature, several broken symmetry phases like charge-density-wave (CDW), 
spin-density-wave (SDW) and spin-Peierls (SP) before becoming superconducting \cite{jer,dumm}. 
The physical mechanism underlying these strongly different orderings 
can be identified as the competition
between lattice distortions and electron dynamics, with a major role played by 
electron interactions. Unfortunately, interaction effects in low dimensional 
electron systems coupled to further degrees of freedom, such as impurities
or phonons, are not fully understood yet. Moreover, in order to describe the
phase diagram of these materials the inclusion of inter-chain coupling 
besides interaction and phonons cannot be avoided. At the very least, 
inter-chain coupling is crucial for the  stabilization of a genuine three 
dimensional long range order, like antiferromagnetism or superconductivity.
However, we believe that, as a first step, a thorough understanding of the behavior of 
purely 1D strongly correlated electron systems coupled to adiabatic phonons
is necessary. 

By now, the physics of one dimensional 
electron systems in a rigid lattice is well understood \cite{oned}:
spin isotropic models display a metallic phase, the celebrated 
Luttinger Liquid (LL), with gapless charge and spin excitations. 
Umklapp processes occurring at commensurate fillings drive
the opening of a charge gap, while attractive interactions
may lead to pairing and then to a spin gap. The presence of a
gapless branch of excitations is intimately related to the 
long range power law behavior of correlations and to the divergence
of selected susceptibilities in these models. Remarkably, in 1D
all these features are governed by a single parameter, denoted 
by $K_\rho$ \cite{oned}. In the LL phase both CDW and SDW susceptibilities 
at wave vector $q=2k_F$ diverge for $K_\rho \le 1$ 
while the CDW susceptibility at $q=4k_F$ is singular for $K_\rho \le 1/2$. 
Eventually, superconducting fluctuations become relevant for $K_\rho \ge 1$.
The $K_\rho$ parameter has been numerically evaluated as a function
of the coupling constants in several 1D models \cite{oned}.
If the Luttinger Liquid is weakly coupled to classical phonons, 
second order perturbation theory predicts the occurrence of
lattice distortions of modulation $q$ whenever the associated CDW
susceptibility diverges. 

In order to substantiate these perturbative results and to extend 
them to finite phonon coupling, the numerical study of simple
lattice models on finite chains may indeed be profitable. The 
repulsive extended Hubbard model has been extensively
investigated \cite{mazdix,maztou,mazclay,rierapoil,kuwa} and the interplay  
between the $2k_F$ and $4k_F$ instabilities has been 
discussed. In such a class of models, the 
antiferromagnetic correlations are generated by virtual 
hopping processes and do not lead to an effective 
attraction between electrons. As a consequence, the key parameter 
$K_\rho$ is restricted to the ``repulsive" range 
$K_\rho \le 1$. By contrast, in the t-J   
model the magnetic coupling $J$ is unrelated to the 
on site repulsion ($U=\infty$) and $K_\rho$ 
acquires values in the extended domain 
$1/2 \le K_\rho < \infty$ \cite{ogata}.
An investigation of the 1D t-J model coupled to
adiabatic phonons would then allow to single out
the role of antiferromagnetic fluctuations in 
driving lattice distortions and stabilizing
CDW, SDW or SP ground states. This is the 
program we intend to pursue in the following sections.

\section{The model}

One of the simplest models able to describe the low energy properties 
of strongly correlated electron systems 
is the t-J model. Here, contrary to the Hubbard model,
the magnetic interaction is explicitly included in the hamiltonian while the 
strong on site Coulomb repulsion acts via the single occupancy constraint. 
The phase diagram of the t-J model in a rigid 1D lattice has been determined \cite{ogata}
and analytic solutions are known in the $J\to 0$ limit (spinless-fermion case)\cite{shiba}
and at $J/t = 2$ (super-symmetric case)\cite{bares,bares1}, besides at half filling, i.e.
for electron density $n=1$, where it reduces to the Heisenberg chain. 
In the latter case the model is known to be unstable toward 
dimerization (spin-Peierls instability) when coupled to an elastic lattice\cite{crfish}. 
As a result, a periodic modulation of the lattice at wave-vector $q=2k_F=\pi$ 
sets in and the system 
displays a bond-order wave (BOW): adjacent spins pair in a singlet state 
and a gap in the spin spectrum opens up. 
When phonons are treated in the adiabatic limit, this transition has been 
predicted for arbitrarily weak spin-lattice coupling\cite{crfish}.

At finite doping, i.e. for electron density $n < 1$, the pure 1D t-J model 
enters the Luttinger Liquid phase 
which extends up to $J/t\lesssim 2.3$ (for $n=1/2$) with growing superconducting
fluctuations until, at larger $J/t$, phase separation sets in.
Contrary to the purely repulsive Hubbard model, the t-J model explicitly exhibits 
the attraction mechanism between electrons mediated by antiferromagnetic fluctuations 
which is likely to be present in real magnetic materials. 

In the following, we study 1D t-J model coupled with classical phonons and we investigate 
by numerical diagonalization technique the interplay between electron interactions 
and lattice distortions. We focus our attention on the competition, or sometimes the cooperative 
behavior, between charge ordering and lattice instabilities and we examine its
consequences on the magnetic properties of the materials. 
The calculations will be performed at quarter filling ($n=1/2$), which 
is the appropriate choice to mimic the behavior of the organic compounds \cite{dumm},
and our results are summarized in Fig. 1 which shows a 
tentative zero temperature phase diagram of the model. 

The one-dimensional t-J model coupled with adiabatically phonons is defined by the 
hamiltonian 
\begin{eqnarray}
H &=& - \sum_{i} {\tonda{1-\delta_i}}
\tonda{\widetilde{c}_{i,\sigma}^{\dagger}\widetilde{c}_{i+1,\sigma}+ h.c.} + \nonumber \\
&+& J\sum_{i}\tonda{1-\delta_i} \tonda{\gr{S}_i\cdot \gr{S}_{i+1}- \frac{1}{4}n_in_{i+1}} + 
\nonumber \\
&+& \frac{1}{2}K_B \sum_{i} \delta_i^{2}
\label{tj}
\end{eqnarray}
Where $\gr{S}_i$ are spin-$1/2$ operators at the site $i$, 
$\widetilde{c}_{i,\sigma}^{\dagger}=c_{i,\sigma}^{\dagger}\tonda{1-n_{i,-\sigma}}$ 
are electron Gutzwiller-projected creation operators and
$n_i=\sum_\sigma c_{i,\sigma}^{\dagger}c_{i,\sigma}$. We have also set the bare hopping integral $t$ 
to unity thereby fixing the energy scale. The hamiltonian depends on the 
classical variables $\delta_i$ which identify 
the bond distortions defined as $\delta_i=\tonda{u_{i+1}-u_{i}}$ where 
$u_i$ is the displacement of ion $i$ with respect to its equilibrium position.  
The last contribution represents the elastic deformation energy and $K_B$ 
is the spring constant. The form of the electron-phonon coupling adopted here 
follows from a linearization of the expected dependence of the coupling
constants on lattice distortions:  
for small displacements $\delta$ of the ion sites, the hopping and exchange integrals vary 
with the distance as a certain power\cite{harr} $\alpha$ which has been estimated 
in the range $6< \alpha < 14$
$$
J_{i,i+1}\tonda{\delta} = J\tonda{\frac{a+u_i}{a+u_{i+1}}}^{\alpha} 
\simeq J\tonda{1-\alpha \frac{\delta}{a}}
$$
where $a$ is the bare lattice spacing. For simplicity we have taken the 
same $\alpha$ for both the hopping and the super-exchange integrals.
We note that the Hamiltonian $\tonda {1}$ is invariant under the 
rescaling $\alpha \rightarrow \lambda \alpha^*$, 
$K_B \rightarrow \lambda^2 K_B^*$, $\delta \rightarrow \delta^* / \lambda$, 
where $\lambda$ is the rescaling factor.
We can then fix one parameter between $\alpha$ and $K_B$\cite{becca} without affecting the
physics of the model. In the hamiltonian (\ref{tj}) we have set $\alpha=1$. 
Clearly, the physical range where the hamiltonian (\ref{tj}) may be used is 
restricted to $|\delta_i| \ll 1$. 

This model is studied at quarter filling $(n=1/2)$ and zero magnetization
by use of the Lanczos method on a finite ring  $16$ sites . Contrary
to previous studies on the Hubbard model, we choose open shell boundary 
conditions, i.e. periodic boundary conditions when the 
number of lattice sites $L$ is an integer multiple of 8, anti-periodic boundary
conditions otherwise. This choice is motivated by the attempt to mimic the
behavior of an infinite chain in a small ring. As already noticed, a key feature of the 
t-J model in the thermodynamic limit is the presence of gapless excitations 
which lead to a diverging CDW susceptibility, and then to lattice distortions.
In closed shell rings, this feature is lost because of the $O(1/L)$ finite size
gap in the kinetic energy spectrum. As a consequence, unrealistically low 
values of the elastic constant $K_B$ are needed in order to stabilize
lattice distortions \cite{rierapoil}. 

The ground state of the hamiltonian (\ref{tj}) is numerically obtained by
Lanczos technique at fixed $\graf{\delta_i}$.
The lattice distortions are then updated in order to minimize the 
total energy $E(\graf{\delta_i})$. This can be accomplished by 
solving the self-consistency equations which follow from the 
Hellmann-Feynman theorem:
\begin{eqnarray}
\frac{\partial E}{\partial\delta_i}&= & \sum_{\sigma} 
\langle \widetilde{c}_{i,\sigma}^{\dagger}
\widetilde{c}_{i+1,\sigma} + h.c. \rangle + J 
\langle S_i\cdot S_{i+1} 
- \frac{1}{4} 
n_in_{i+1} \rangle 
\nonumber \\
&+& K_B\delta_i + \frac{1}{N} \sum_{j,\sigma} 
\langle \widetilde{c}_{j,\sigma}^{\dagger}\widetilde{c}_{j+1,\sigma} 
+ h.c. \rangle 
- \nonumber\\ 
&-& \frac{J}{N} \sum_j \langle S_j\cdot S_{j+1} - \frac{1}{4} 
n_jn_{j+1} \rangle =0  
\label{self}
\end{eqnarray}
The constraint $\sum_i \delta_i=0$ which reflects the chosen ring geometry 
has been implemented by use of a Lagrange multiplier. 
Here $\langle ...\rangle$ is the ground-state expectation value.  
In order to find the bond length configuration which minimizes
the total energy at given coupling constants $(J, K_B)$ 
we iterate the set of equations (\ref{self}) until convergence is reached. 

\section{Numerical Results}

Here we present the numerical results on the t-J model coupled to adiabatic phonons at quarter filling and
vanishing magnetization for a $16$ site ring. 
In this case  $k_F=n\pi/2 = \pi/4$ and the two expected, competing periodicities 
correspond to tetramerization ($q=2k_F$) and dimerization ($q=4k_F$).
In the former case two electrons belong to the unit cell and then we expect
a paramagnetic insulator behavior with both charge and spin gap. This regime
precisely corresponds to the spin-Peierls phase experimentally found in the 
Fabre salt series. 
Instead, in the presence of lattice dimerization, the model should be a paramagnetic
metal, having a single electron per unit cell. However, the residual Coulomb repulsion
drives the system toward a Mott insulator phase. 
Such a circumstance can be explicitly checked in the 
$J\to 0$ limit, where some exact statement can be made. 

In one dimension, when the kinetic term alone is present, 
the charge degrees of freedom are described by a free spinless fermion gas 
\cite{shiba} whose properties, even in the presence of (adiabatic) phonon coupling, 
can be explicitly derived in the thermodynamic limit. Lattice 
dimerization is stabilized for arbitrarily 
large $K_B$ \cite{qualcuno} but such a symmetry breaking 
cannot be accompanied by a CDW of the same periodicity because reflection across a 
bond leaves the distortion pattern unaltered while interchanging the two sublattices. 
As a result, in the t-J model at $n=1/2$ and $J\to 0$, the charge distribution remains uniform
along the chain while a charge gap opens due to the doubling of the unit cell.
The spin degrees of freedom arrange as in the ground state of the Heisenberg chain 
and the spin correlations can be expressed in a factorized form \cite{parola}:
\begin{equation}
\langle \gr {S}_r\cdot \gr {S}_0 \rangle = 
\sum_{j=2}^{r+1}P^r_{SF}\tonda{j}S_H\tonda{j-1}
\label{spin}
\end{equation}
where $P^r_{SF}\tonda{j}$ is the probability of finding $j$ particles in $\tonda{0,r}$ 
with one particle in $0$ and another in $r$, evaluated in the ground state of the 
dimerized spinless Fermi gas. If we set $N_r=\sum_{i=0}^r n_i$, the probability
$P^r_{SF}\tonda{j}$ can be formally expressed as
$
P^r_{SF}\tonda{j} = \langle n_on_r\delta\tonda{N_r - j} \rangle
$
and $S_H(j)=\langle \gr {S}_j \cdot \gr {S}_0 \rangle_H$ 
is the spin-spin correlation function of the isotropic Heisenberg chain.  
Following Ref. \cite{parola} the asymptotic decay of Eq. (\ref{spin}) 
in the presence of lattice dimerization can be evaluated as
\begin{equation}
\langle \gr {S}_r \cdot \gr {S}_0 \rangle \propto 
\frac{\cos(2k_Fr-\frac{\pi}{4})}{r}
\label{heis}
\end{equation}
much slower than in the standard t-J model at the same electron density.
This asymptotic result valid for any lattice dimerization, can be understood
from Eq. (\ref{spin}) in the simpler limit of decoupled dimers, 
i.e for $\delta_i \sim (-1)^{i+1}$ where 
\begin{equation}
P^r_{SF}(j)=\frac{1}{4}\cdot\cases{ 0 & for $r=1$\cr 
\delta_{j,(r+1)/2} & for $r$ odd $\ne 1$  \cr
\delta_{j,(r+2)/2} & for $r$ even  \cr }
\label{psf}
\end{equation}
Remarkably, Eq. (\ref{heis}) reproduces the same $1/r$ decay of spin 
correlations as in the usual Heisenberg chain.
This form of the magnetic correlations, together with the presence of a 
charge gap shows that the model is indeed in a Mott insulator phase. 
Such a result suggests that lattice dimerization strongly favors antiferromagnetism, 
which will be eventually stabilized by inter-chain coupling giving rise to 
three dimensional antiferromagnetic ordering. Therefore, in the 
$J\to 0$ limit, the t-J model coupled to adiabatic phonons reproduces
one of the phases experimentally found in the quasi one dimensional Fabre 
salts. Now we want to investigate how the enhancement of antiferromagnetic 
coupling induced by the $J$ term modifies this picture. 

The numerical results are summarized in
Fig. 1 where the phase diagram emerging from our analysis is sketched. 
\begin{figure}
\includegraphics[width=0.45\textwidth]{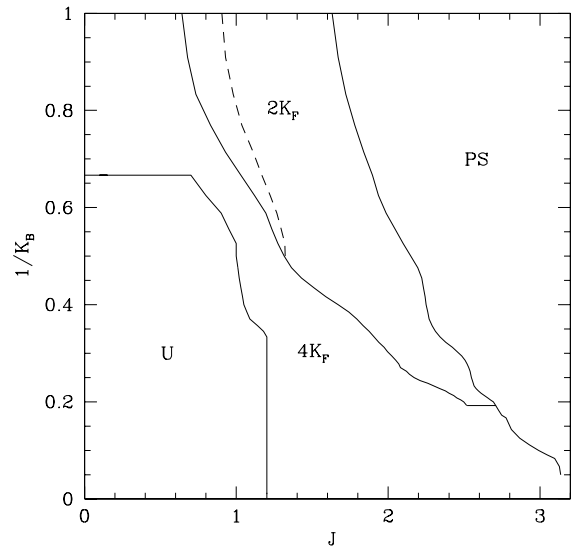}
\caption{\label{phase}
Phase diagram of the t-J model coupled to adiabatic phonons (\ref{tj})
obtained by numerical diagonalization of a $16$ site chain. The boundaries 
of the uniform phase ($U$), the dimerized BOW ($4k_F$), the tetramerized 
BCDW ($2k_F$) and the phase separated region are shown by full lines. The
dashed line represents a crossover between the bond pattern shown in Figs.
2c (left) and 2d (right).
}
\end{figure}
No lattice or spin symmetry has been imposed and the Lanczos diagonalization
has been performed in the full Hilbert space of the model. This strongly 
limits the feasible sizes to the range $L \le 16$. In all cases the ground state
we found shows definite lattice periodicity, the result does not depend
on the starting bond configuration or on the convergence requirements. 
Noticeable exceptions to this statement are the states in the phase 
separated region (PS in Fig. 1) at very large $J$,
where the energy landscape presents several local minima. 
In order to check the stability of the numerical results, we performed 
a set of Lanczos diagonalizations in the reduced Hilbert space 
obtained by selecting a given unit cell (e.g. two sites or four sites).
We always found consistency between the numerical output in the
full and in the reduced Hilbert space except in the PS 
region, where the ground state energy obtained in {\sl every} subspace 
of definite translational symmetry is larger than the ground state 
energy found in the full Hilbert space. This feature 
identifies the boundary of the PS region. 
We will not discuss this unphysical regime any further.

The distortion patterns we found correspond to periodicities of growing 
wavelength moving from small $J$ to larger $J$:

\begin{itemize}
\item[-]  uniform (U in Fig. 1) 
\item[-]  dimerized ($4k_F$ in Fig. 1) 
\item[-]  tetramerized ($2k_F$ in Fig. 1)
\end{itemize}

A general feature of the phase diagram we obtained is the 
rather weak dependence of the phase boundaries on the spring constant
$K_B$: this shows that the antiferromagnetic coupling $J$ is indeed 
the driving force which selects the periodicity of lattice distortions.
In the uniform phase, $\delta_i=0$ and the system is a metal according to the 
theory of Luttinger Liquids.
The numerical data predict a uniform phase also in the
previously discussed $J\to 0$ and large $K_B$ limit where the charge degrees of 
freedom behave as spinless fermions which are known to undergo
spontaneous dimerization when coupled to adiabatic phonons. 
This shows that, at least in this region, strong finite size effects are 
present in our small cluster. In order to investigate this problem,
we performed few Lanczos diagonalizations 
for the $L=20$ site chain in the Hilbert sub-space defined by the eigenstates 
of the four site translation operator. This calculation allows to
study the interplay between uniform solution, dimerization and tetramerization
in this cluster. 
The phase boundary between the uniform and the dimerized phase 
shrinks from $J\sim 1.2$ to $J\sim 0.8$ at $K_B=3$ while the
onset of tetramerization hardly moves. Therefore we are led to conclude that, 
as already discussed in previous numerical studies of
the Hubbard model \cite{mazclay}, the very existence of the uniform phase 
in the adiabatic limit is likely to be an artifact of the smallness of the 
cluster considered, while the other phase are much less affected by finite size 
problems. 
\begin{figure}
\includegraphics[width=0.4\textwidth]{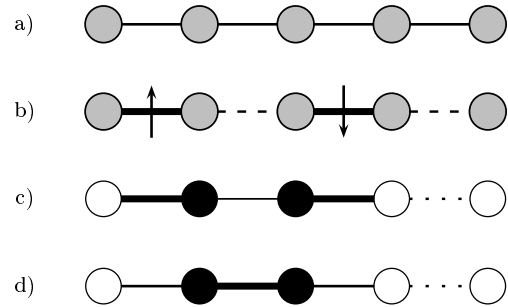}
\caption{\label{instab}
Pictures of the competing density waves in the ground state. a) Uniform state (U), 
b) $4k_F$ BOW and uniform charge, c) $2k_F$ BCDW tetramerized phase ($S-W-S-W'$),
d) $2k_F$ BCDW tetramerized phase ($U-S-U-W$). 
}
\end{figure}

At larger $J$ the dimerized $4k_F$ state prevails and according to the 
previous analysis the chain behaves as a Mott insulator. 
While the electron density remains uniform, the lattice develops a 
bond ordered wave (BOW): a charge gap opens due to the Umklapp processes triggered 
by the doubling of the unit cell and the cell-cell spin correlations
show an extremely slow decay characterized by a diverging antiferromagnetic
susceptibility. When three dimensional coupling is allowed for, a genuine
antiferromagnetic long range order sets in. 

Finally, by further increasing
$J$, the lattice periodicity changes and the unit cell doubles again giving rise
to lattice tetramerization. In this case, a weak $2k_F=\pi/2$ charge density wave 
develops on top of the lattice distortion giving rise to a 
bond/charge density wave (BCDW) \cite{maztou,mazclay}: 
\begin{eqnarray}
\langle n_i \rangle &=& \frac{1}{2} + A_{2k_F}\cos{\tonda{\frac{\pi}{2} i+\frac{\pi}{4}}}\nonumber\\
\delta_i&=&B_{4k_F}\cos{\tonda{\pi i}} +B_{2k_F}\cos{\tonda{\frac{\pi}{2}i}}
\label{dist}
\end{eqnarray}
The phase shifts shown in Eq.(\ref{dist}) are those obtained by the numerical diagonalization
and correspond to a CDW of the type $-++-$, where $+$ ($-$) identifies a 
local density higher (lower) than the average ($n=1/2$). 
No significant $4k_F$ component in the CDW is detected in the BCDW,
in agreement with previous investigations of the Hubbard model. 
Interestingly, the bond pattern
depends on the relative magnitude of the $B_{4k_F}$ and $B_{2k_F}$ amplitudes:
For $B_{4k_F} > |B_{2k_F}|$ the ordering is of the type $S-W-S-W'$ (Fig. 2c)
while for $|B_{4k_F}| < B_{2k_F}$ the sequence is $U-S-U-W$ (Fig. 2d). Here, $W$, $U$
and $S$ respectively represent a weak ($W$) almost undistorted ($U$) and strong ($S$) bond. 
Although no real transition, but rather a crossover, is present between these
two cases, we can identify a small region (shown in Fig. 1) where the bond ordering
is of the first type, while in the remaining region of stability of the tetramerized 
phase the bond lengths arrange as in Fig. 2d. The crossover between these two 
different realizations of the $2k_F$ phase may be understood by noting that 
the $S-W-S-W'$ BOW (Fig. 2c) appears as an intermediate regime between the $4k_F$ 
region and the stable $U-S-U-W$ (Fig. 2d) ordering, which indeed dominates in a large
portion of the phase diagram. 

Notice that previous studies 
of the extended Hubbard model \cite{mazclay,rierapoil} found only a pattern of the first type,
which is often referred to as the dimerization of an already dimerized state. 
In such a regime, our calculations suggest the occurrence of a transition
between a low temperature tetramerized phase and dimerization at higher temperature.
In order to investigate this possibility, which is indeed experimentally 
found in $DI-DCNQI$ compounds, we evaluated the ground state energy of the 
16 site ring at given lattice distortion: We kept the amplitude of the
two strong bonds  fixed while varying the difference $\Delta$ between the two weak bonds 
of the unit cell $S-W-S-W'$. The result, shown in Fig. 3 for $J=0.7$ and $K_B=1$,
displays a double well structure characterized by a small
energy barrier between the tetramerized ground state at $\Delta\sim\pm 0.06$ and the
dimerized configuration corresponding to $\Delta=0$.

\begin{figure}
\includegraphics[width=0.45\textwidth]{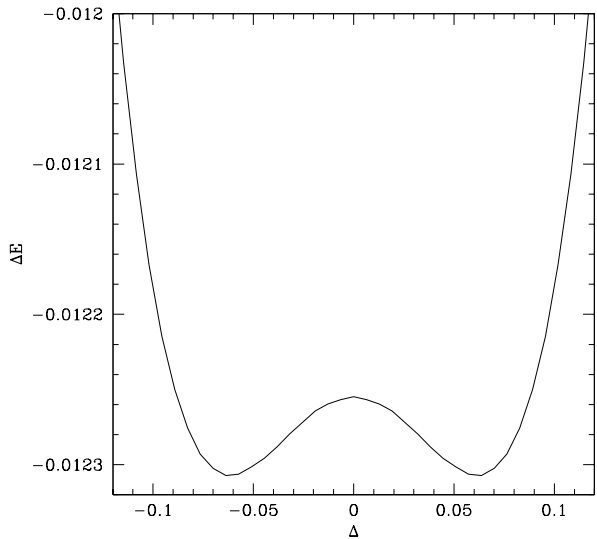}
\caption{\label{land}
Energy landscape at $J=0.7$ and $K_B=1$. The ground state energy,
referred to the undistorted state, is plotted as a function
of the difference $\Delta$ between the two weak bonds in the 
$S-W-S-W'$ phase. The strong bond is fixed at the equilibrium value 
$\delta=0.15925$ }
\end{figure}

Whenever lattice tetramerization prevails, two electrons belong to the 
unit cell and then the chain becomes a paramagnetic band insulator 
as in the spin-Peierls phase. We believe that, because of the presence of 
both charge and spin gap, this phase is stable toward 3D coupling. 

It is interesting to investigate the energy difference between the 
electronic states corresponding to the low energy lattice
distortions which are solutions of Eq. (\ref{self}). 
Such a calculation may provide some useful information 
about the temperature range where an ordered phase is indeed stable. 
In Fig. 4 we plot the energy of different distortions as a function of $J$ at fixed $K_B=3$.  
More precisely, we plot the energy difference between the lowest
metastable state of a given periodicity ($4k_F$ or $2k_F$) and the
uniform solution. 
\begin{figure}
\includegraphics[width=0.48\textwidth]{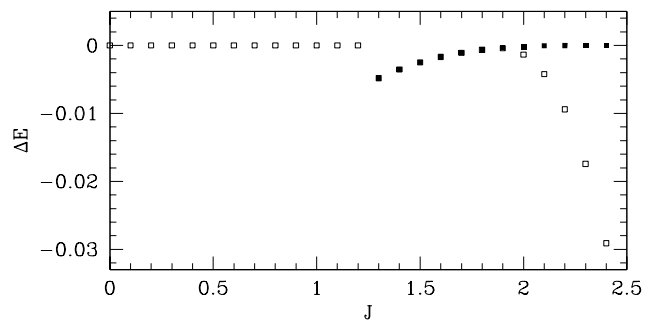}
\caption{\label{gain}
Energy gains between different distortions at fixed $K_B=3$ in a $L=16$ site ring.
Open (full) squares refer to a four (two) site unit cell.
}
\end{figure}

The numerical data also suggest first order boundaries between the 
uniform and the dimerized phase,
as can be inferred by the sudden jump of the appropriate 
Fourier component of the lattice distortion $\{\delta_i\}$ shown in 
Fig. 5 for a couple of values of $K_B$. 
Our results do not allow to draw any definite conclusion about the order of the
transition between the dimerized and the tetramerized phase. 
\begin{figure}
\includegraphics[width=0.45\textwidth]{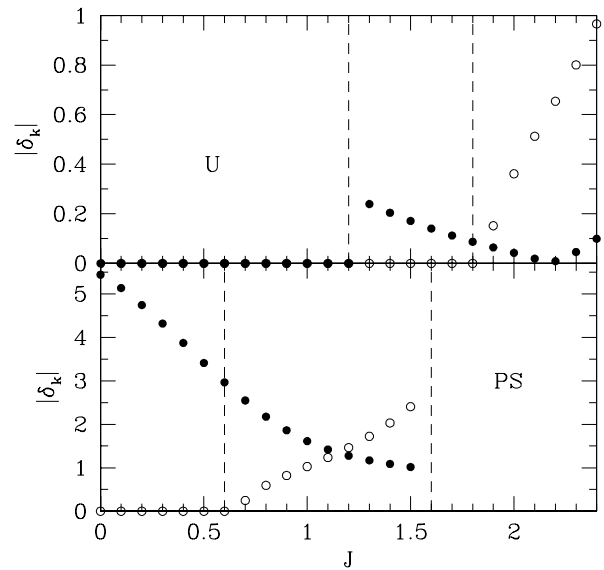}
\caption{\label{fourier}
Amplitude of the $4k_F$ (full dots) and $2k_F$ (open dots) 
Fourier components of the bond distortions in the ground state 
of the model. Upper panel: $K_B=3$, lower panel $K_B=1$ }
\end{figure}

Although the periodicities of the lattice distortions we found 
numerically precisely correspond to those expected on the basis of
the instabilities of the homogeneous t-J model, some
discrepancy on the sequence of distortion patterns does occur. 
As previously discussed, the t-J model is characterized by 
a $K_\rho$ parameter always larger than $1/2$ in the whole
range $J>0$. This would imply that, at least in the perturbative
regime of small distortions (i.e. for $K_B\to\infty$), pure dimerization 
should never be observed. Instead, tetramerization should be seen 
for $0 < J \lesssim 2.3$ which, according to Ref. \cite{ogata}, corresponds
to $K_\rho \le 1$. While the upper limit for $2k_F$ distortions roughly agrees
with the Lanczos diagonalization data, the numerical results clearly show 
that, for large $K_B$, dimerization and not tetramerization provides the
most stable distortion pattern. This behavior can be understood by noting that
the {\sl amplitude} of the $2k_F$ singularity in the charge response of
the uniform t-J model is strongly suppressed for $J \lesssim 2$, as
already noted in Refs.\cite{assaad,troyer}. This suggests that even if
in the thermodynamic limit a $2k_F$ singularity may eventually prevail in the
whole phase diagram, at small $J$ a clear {\sl dimerization} 
pattern with only weak tetramerization corrections will be visible. 

\section{Conclusions}

In summary, we determined the phase diagram of the 1D t-J model
adiabatically coupled to the lattice by use of Lanczos diagonalization
in small clusters. We found, for increasing antiferromagnetic coupling,
a metallic uniform phase, a BOW phase which displays the typical features
of a Mott insulator and preludes to 3D antiferromagnetic (AF) order, a BCDW phase
which realizes the spin-Peierls scenario. Therefore, by {\sl increasing} the 
AF coupling $J$ in the model the systems goes from antiferromagnetic to 
paramagnetic because of the change in distortion pattern determined
by the charge degrees of freedom. This shows that in the t-J model charge and 
spin ordering compete due to the coupling to the lattice and strong 
super-exchange interaction may lead to a spin-Peierls ground state. 
Remarkably, the phases we found are 
all present in the typical zero temperature phase diagram of the quasi 1D 
organic materials \cite{jer}. Moreover, these phases do indeed occur 
in the same sequence as we found when uniform pressure is applied to 
the material. The effect of pressure is mainly to increase the hopping 
amplitude (which we set as unit of energy) and then our results are 
compatible with the experiments provided the effective intra-chain 
superexchange coupling $J$ is only weakly dependent on the applied 
pressure. If this is the case, the t-J model may indeed represent
a useful theoretical framework for the interpretation of the behavior
of the quasi 1D organic materials. 

The effects of charge ordering in this class of materials has been mainly investigated by 
using an extended Hubbard model coupled with the lattice \cite{mazclay,rierapoil}.
A phase diagram, of the ground state, obtained as a function of the electron-phonon 
couplings show the same periodicities that we found with the exception of
the one in Fig. 2d. Notice that the strong coupling limit of the
extended Hubbard model is related to an effective t-J model \cite{brink} by 
$J=\frac{4t^2}{U-V}$, which restricts the mapping to the $J \ll 1$ regime.

It is a pleasure to thank F. Becca for stimulating discussions.
Financial support by the MIUR-PRIN program is gratefully acknowledged.

\end{document}